\documentclass[aps,prl,showpacs,twocolumn,superscriptaddress,floatfix]{revtex4}

\usepackage{graphicx}% Include figure files
\usepackage{dcolumn}% Align table columns on decimal point
\usepackage{bm}% bold math
\usepackage{amsmath}
\usepackage{amssymb}

\begin{document}

\renewcommand{\vec}[1]{\mbox{\boldmath $#1$}}

%\preprint{}

\title{Molecular spin-liquid state in spin-3/2 frustrated spinel HgCr$_2$O$_4$}


\author{K. Tomiyasu}
\email[Electronic address: ]{tomiyasu@m.tohoku.ac.jp}
\affiliation{Department of Physics, Tohoku University,
Aoba, Sendai 980-8578, Japan}
\author{H. Ueda}
\affiliation{Division of Chemistry, Graduate School of Science, Kyoto University,
Sakyo, Kyoto 606-8502, Japan}
\author{M. Matsuda}
\affiliation{Neutron Scattering Science Division, Oak Ridge National Laboratory,
Oak Ridge, Tennessee 37831, USA}
\author{M. Yokoyama}
\affiliation{Faculty of Science, Ibaraki University,
Mito, Ibaraki 310-8512, Japan}
\author{K. Iwasa}
\affiliation{Department of Physics, Tohoku University,
Aoba, Sendai 980-8578, Japan}
\author{K. Yamada}
\affiliation{WPI AIMR, Tohoku University,
Aoba, Sendai 980-8577, Japan}
%\thanks{}


\date{\today}

\begin{abstract}
A hexamer-type spin excitation seen in spinel chromates $A$Cr$_2$O$_4$ ($A$ = Mg, Zn, or Cd) is the representative spin-liquid-like state caused by geometrical frustration. To clarify an origin of the state, we comparatively studied spin excitations in an isomorphic material HgCr$_2$O$_4$ by inelastic neutron scattering, and observed a different molecular-type excitation. 
Numerical analyses performed using model Hamiltonians suggest that these two types of spin excitations originate from a spin-3/2 molecular singlet hidden in a magnetically ordered phase. The difference between the molecular types is explained by the difference in the kind of exchange interactions occurring in the chromates.
\end{abstract}

%\keywords{geometrical frustration, inelastic neutron scattering, spin molecule, dimer}

\maketitle

%
%\section{Introduction}
%
%Geometrical frustration
Remarkable developments have been made in the fields of solid-state physics in magnetic systems and strongly correlated electron systems using the concept of geometrical spin frustration. In frustrated systems, not all classical-spin pairs can be arranged antiferromagnetically on a triangle lattice and a tetrahedral lattice~\cite{Pauling_1935,Wannier_1950,Anderson_1956}. Therefore, frustration suppresses magnetic ordering and promotes spin-liquid-like fluctuations (zero-energy excitations) in a low-temperature paramagnetic phase, e.g., spin molecules, spin ices, and spin vortices~\cite{Ballou_1996,Lee_2002,Bramwell_2001,Kanada_2002,Kawamura_2010}. Recently, dynamical spin molecules were found to exist as non-dispersive gapped excitation modes within a magnetically ordered phase, where frustration was assumed to be relieved by a lattice distortion~\cite{Tomiyasu_2008}.

%Spin molecule
Formation of a spin molecule is one of the representative effects of geometrical frustration in a system~\cite{Lee_2002}. A spin molecule refers to a spin cluster that is spatially confined within a geometrical shape such as an atomic molecule. Intermolecular correlation among spin molecules is negligible in comparison to the intramolecular correlation; this behavior is similar to the quasiparticle approximation in Landau's Fermi liquid theory~\cite{Tomiyasu_2008}. For example, a paramagnetic phase in the spinel $A$Cr$_2$O$_4$ (nonmagnetic $A$ = Mg, Zn, Cd) is known to undergo non-frustrated hexamer-type spin fluctuations so as to avoid frustration (hexa-I in Fig.~\ref{fig:models}(b))~\cite{Suzuki_2007,Tomiyasu_2008,Lee_2002,Chung_2005}. The magnetic ions Cr$^{3+}$ ($(t_{2g})^3$, spin $S=3/2$) form a corner-sharing tetrahedral lattice called a pyrochlore lattice; such a lattice has kagome and triangle planes stacked alternately along the [111] direction and is geometrically frustrated.

%HgCr2O4: common features
HgCr$_2$O$_4$ is isomorphic to the aforementioned spinel chromates. Its Curie-Weiss temperature $\Theta_{CW}$ has been estimated to be $-32$ K, indicating an antiferromagnetic spin correlation~\cite{Ueda_2006}. It undergoes antiferromagnetic long-range ordering with propagation vectors (1/2,0,1) and (1,0,0) below $T_{N}=6$ K, which is much lower than $\Theta_{CW}$ owing to frustration~\cite{Ueda_2006,Matsuda_2007}. A complex magnetic structure of HgCr$_2$O$_4$ determined by powder neutron diffraction also suggests that antiferromagnetic first-neighbor exchange interactions govern the spin system, as is the case with the other chromates~\cite{Matsuda_2007}.

%HgCr2O4: different features
However, the magnitude of $\Theta_{CW}$ and the degree of frustration $|\Theta_{CW}/T_{N}|\simeq5$ are both lower than those of the other chromates; for example, the $\Theta_{CW}$ values of Mg, Zn, and Cd chromates are about $-370$, $-390$, and $-70$ K, and their corresponding $|\Theta_{CW}/T_{N}|$ values are about $30$, $31$, and $9$, respectively~\cite{Ueda_2005,Ueda_2006}. A low $\Theta_{CW}$ indicates a scaling-down of exchange interactions. A low $|\Theta_{CW}/T_{N}|$, in turn, indicates a suppression of frustration by relatively strong further-neighbor exchange interactions and spin-lattice coupling, which also explain the wide magnetization plateau observed under a high magnetic field from 10 to 27 T~\cite{Penc_2004,Motome_2006,Tanaka_2007,Matsuda_2007}.

Thus, HgCr$_2$O$_4$ might exhibit spin excitations other than hexa-I. To the best of our knowledge, no inelastic neutron scattering experiments on HgCr$_2$O$_4$ have yet been performed, probably because of the strong neutron absorption of Hg nuclei.
%~\cite{ScattL_1992}.

%This work
In this study, therefore, we performed powder inelastic neutron scattering experiments on HgCr$_2$O$_4$ above and below $T_N$ without a magnetic field. This Letter reports the discovery of another type of dynamical spin molecules from the experiments. It also discusses the origin of the difference between molecular types and proposes a quantum-mechanical picture for spin molecules by numerical analyses using model Hamiltonians.

%
%\section{Experiments}
%
A powder sample of HgCr$_2$O$_4$ was synthesized by thermal decomposition of Hg$_2$CrO$_4$ in an evacuated silica tube~\cite{Ueda_2006}. Some preliminary neutron experiments were performed on the triple-axis spectrometer TAS-2 of the Japan Atomic Energy Agency (JAEA), installed in the thermal guide tube of the JRR-3 reactor in JAEA. Inelastic neutron scattering experiments were performed on the triple-axis spectrometer HER (C1-1) at the Institute for Solid State Physics (ISSP), University of Tokyo; this spectrometer is installed in the cold guide tube of the same reactor. The energy of the final neutrons, $E_f$, was fixed to 3.6 meV with a horizontal collimation sequence of guide-open-radial-open, where the radial collimator has 3 blank channels. A horizontal focusing analyzer merged the scattered neutrons in a range of about 7 degrees scattering of angle. A cooled Be filter and a pyrolytic graphite Bragg-reflection filter efficiently eliminated the half-lambda contamination.
%for the incident neutron energy $E_{i}<5.1$ meV and for $E_{i}>5.1$ meV, respectively.
The powder (4.5 g) was filled in a thin aluminum foil, which was then shaped into a hollow cylinder of diameter 25 mm in order to minimize the strong neutron absorption. Then, the sample was enclosed in an aluminum container with $^4$He exchange gas, which was placed under the cold head of a closed-cycle $^4$He refrigerator.

%
%\section{Results}
%

\begin{figure}[htbp]
\begin{center}
\includegraphics[width=3.4in, keepaspectratio]{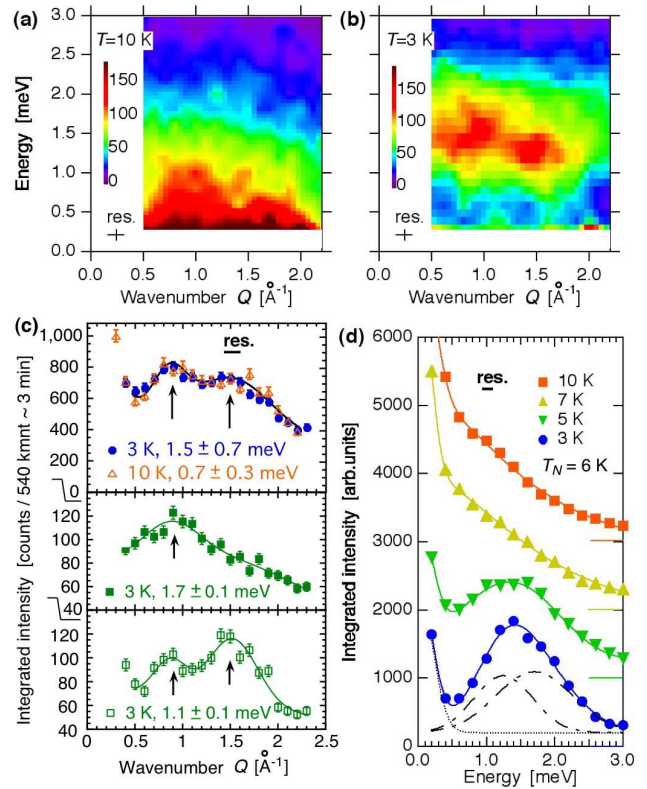}
\end{center}
\caption{\label{fig:data} (Color online) Measured inelastic neutron scattering data for powder HgCr$_2$O$_4$. (a)(b) Scattering intensity distributions in $(Q,E)$ space above and below $T_N$, respectively. The vertical tone indicates the scattering intensity in arbitrary units. (c) $Q$ dependence of $E$-integrated intensity. The integration ranges are given in each panel.
%$1.7\pm0.1$ meV in (b) in the top panel, $1.1\pm0.1$ meV in (b) in the middle panel, and $1.5\pm0.7$ meV in (b) and $0.7\pm0.3$ meV in (a) in the bottom panel.
The solid lines are a guide to the eye. (d) $E$ dependence of intensity integrated from $Q=0.5$ to 2.0 {\AA}$^{-1}$ measured at several temperatures. In the data for 10, 7, and 5 K, the vertical origins are shifted to the height shown by horizontal solid lines on the right side, and the solid curves are a guide to the eye. The solid curve for the 3 K data is obtained by a multi-Gaussian fitting, indicated by the dotted and broken lines. The errors are smaller than the symbols used.
%(e) Temperature dependence of elastic peak intensity at the scattering angle 59.3$^{\circ}$ corresponding to 3/2 0 1 magnetic reflection.
}
\end{figure}

\begin{figure}[htbp]
\begin{center}
\includegraphics[width=3.0in, keepaspectratio]{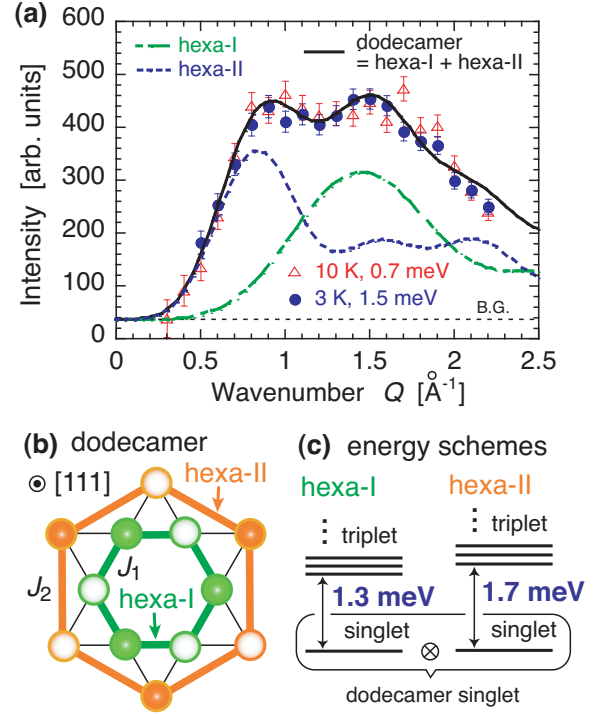}
\end{center}
\caption{\label{fig:models} (Color online) Dynamical structure modeling of spin-excitation modes in HgCr$_2$O$_4$. (a) Experimental $Q$ dependence (symbols) and calculated $Q$ dependence (curves). The experimental data are obtained by subtracting a low-$Q$ background component fitted with a Lorentzian from the raw data (top panel in Fig.~\ref{fig:data}(c)). The latter curves are obtained from a kagome-star dodecamer model consisting of hexa-I and hexa-II (b), and show the common results calculated independently using Eqs.~(\ref{eq:xs_el}) and (\ref{eq:xs_general}) (see the text). The broken and dotted curves broaden approximately with reciprocals of the hexa-I and hexa-II spatial sizes around the main peaks at $Q\simeq1.5$ and 0.9 {\AA}$^{-1}$, respectively. (b) Kagome-star dodecamer model. The subunits hexa-I and hexa-II are depicted by bold lines. The closed and open circles indicate up and down spins, respectively, fluctuating dynamically in arbitrary directions. Exchange interactions $J_1$ and $J_2$ are also defined. (c) Energy schemes of hexa-I and hexa-II obtained by exact diagonalization of Eqs.~(\ref{eq:H_hexa1}) and (\ref{eq:H_hexa2}). The hexa-I part in HgCr$_2$O$_4$ is converted into hexa-I in the other chromates by tuning the energy scale~\cite{cmmnt_1}. }
\end{figure}

%overall S(Q,E) and Q dependence
Figures~\ref{fig:data}(a) and \ref{fig:data}(b) show the scattering intensity distributions in momentum $Q$ and energy $E$ space, measured above and below $T_N$, respectively. Above $T_N$, quasielastic scattering with two peaks at $Q\simeq0.9$ and 1.5 {\AA}$^{-1}$ is observed, as shown in Fig.~\ref{fig:data}(a). In contrast, below $T_N$, this double-$Q$ scattering becomes gapped resonance-like excitations around 1.5 meV, as shown in Fig.~\ref{fig:data}(b). The central values of resonance energies at the two $Q$'s seem to be slightly different.

The $Q$ dependence of $E$-integrated intensity with higher statistics is shown in the top panel of Fig.~\ref{fig:data}(c). A homothetic double-$Q$ curve is observed at the two temperatures. The peak around 1.5 {\AA}$^{-1}$ is attributable to hexa-I, since the $Q$ value coincides with that observed for MgCr$_2$O$_4$ and ZnCr$_2$O$_4$~\cite{Tomiyasu_2008,Lee_2000}. The emergence of the peak at around 0.9 {\AA}$^{-1}$ is characteristic of HgCr$_2$O$_4$.

%

%E dependence
In order to confirm that the resonance-like excitations consist of two modes with different $Q$ peak positions and similar excitation energies, we compared the $Q$ dependences of intensities integrated in two narrow $E$ ranges (see the middle and bottom panels in Fig.~\ref{fig:data}(c)). The 1.7 meV data show only the single peak around 0.9 {\AA}$^{-1}$ (middle), whereas the 1.1 meV data show the double-$Q$ structure with relatively stronger intensity around 1.5 {\AA}$^{-1}$ (bottom), evidencing the coexistence of the two modes.

Figure~\ref{fig:data}(d) shows $E$ spectra integrated from $Q=0.5$ to 2.0 {\AA}$^{-1}$ and measured at several temperatures around $T_N$. With decreasing temperature, the shape of the quasielastic spectra observed above $T_N$ changes abruptly to the gapped shape below $T_N$, similar to a first-order transition. The spectral profile appreciably becomes asymmetric at 3 K. A double-Gaussian fitting with a constant background and an elastic line gives central values of 1.3 and 1.7 meV for the two modes.

%Summary of results
To summarize, we discovered two almost degenerated spin-excitation modes below $T_N$: the first mode is centered around 1.3 meV and 1.5 {\AA}$^{-1}$ and is most probably based on hexa-I, and the second mode is centered around 1.7 meV and 0.9 {\AA}$^{-1}$. The modes are quasielastic above $T_N$.

%
%\section{Analyses}
%

%QENS
Next, in order to determine the dynamical spin structures, we analyzed the $Q$ dependence of quasielastic excitations, though it is difficult without single-crystal data. The cross section of spin molecules is described by
\begin{equation}
S(Q) = C_{1} |F(Q)|^2 \Biggl(\bigl| \sum_{j=1}^{N} S_{j} \exp({i \vec{Q} \cdot \vec{r}_{j}}) \bigr|^2 \Biggr) + {\rm B.G.},
\label{eq:xs_el}
\end{equation}
where $C_{1}$ is a scale factor of the experimental scattering intensity,
${\rm B.G.}$ is a constant background,
$F(Q)$ is the magnetic form factor of Cr$^{3+}$, for which the Watson-Freeman form factor was used below~\cite{Watson_1961},
$j$ denotes the site of Cr$^{3+}$,
$N$ is the total number of sites in a molecule, 
$\vec{r}_{j}$ is the position, and 
$S_j$ takes only $\pm1$ corresponding to collinear spins that dynamically fluctuate in arbitrary directions~\cite{Tomiyasu_2008,Lee_2002}. The parentheses indicate an orientational average with resolution convolution.
Using Eq.~(\ref{eq:xs_el}), we determined a kagome-star dodecamer model, which can be regarded as a superposition of the first-neighbor hexamer (hexa-I) and a second-neighbor hexamer (hexa-II), as shown in Fig.~\ref{fig:models}(b). The molecules are formed on the kagome planes. Figure~\ref{fig:models}(a) shows the calculated $Q$ dependence of intensity for the dodecamer (solid curve), which is in good agreement with the experimental data.

We also note that the dodecamer model cannot be definitively distinguished from the model of the superposition of individual hexa-I and hexa-II. Figure~\ref{fig:models}(a) also shows the calculated curves for hexa-I (broken) and hexa-II (dotted), whose summation with equal weight gives the same curve as the solid one. Hereafter, we refer to both these models as the dodecamer model without any distinction.

%

%Model Hamiltonian and basis states for inelastic scattering
Using the dodecamer model, we analyzed the resonance-like excitations observed below $T_N$ by assuming the following effective Hamiltonians:
\begin{eqnarray}
\label{eq:H_hexa1}
\hat{H}_{\rm hexa\mathchar`-I}&=&J_{1}\sum_{\langle i,j \rangle{\rm \mathchar`-I}} \hat{\vec{S}_{i}} \cdot \hat{\vec{S}_{j}}, \\
\label{eq:H_hexa2}
\hat{H}_{\rm hexa\mathchar`-II}&=&J_{2}\sum_{\langle i,j \rangle{\rm \mathchar`-II}} \hat{\vec{S}_{i}} \cdot \hat{\vec{S}_{j}},
\end{eqnarray}
where $S_{i}=3/2$, $i$ and $j$ denote the hexagonal sites 1 to 6, $\sum_{\langle i,j \rangle{\rm \mathchar`-I}}$ and $\sum_{\langle i,j \rangle{\rm \mathchar`-II}}$ denote summation over all first-neighbor and second-neighbor $\vec{S}$ pairs (not doubly counted), and $J_1$ and $J_2$ denote the first- and second-neighbor exchange interactions that are antiferromagnetic, as expected from the quasielastic excitations (Fig.~\ref{fig:models}(b)). We used 4096 ($=4^6$) basis states of $|S_{1}^{z},S_{2}^{z},S_{3}^{z},S_{4}^{z},S_{5}^{z},S_{6}^{z}\rangle$, where $S_{i}^{z}=\pm3/2$ and $\pm1/2$. 
%We avoided direct treatment of the 12-site Hamiltonian with a large number of basis states ($4^{12}\sim10^{7}$).
After obtaining molecular ground states $\mid\lambda_{0}\rangle$ and $n$th excited states $\mid\lambda_{n}\rangle$ with excitation energy $E_n$ by exact diagonalization of Eqs.~(\ref{eq:H_hexa1}) and (\ref{eq:H_hexa2}), the cross section of magnetic inelastic neutron scattering at $E=E_n$ can be calculated by
\begin{multline}
S(Q,E_n) =  %\\
C_{2} |F(Q)|^2 \delta(\hbar\omega-E_n)
\Biggl( \sum_{\alpha,\beta=1}^{3} (\delta_{\alpha\beta} - \frac{Q_{\alpha}Q_{\beta}}{|\vec{Q}|^2}) \\ %\phantom{0} \times \\
\sum_{j,j^{\prime}=1}^{N}
\langle\lambda_{0}\mid \hat{S}_{j}^{\alpha} \mid\lambda_{n}\rangle
\langle\lambda_{n}\mid \hat{S}_{j^{\prime}}^{\beta} \mid\lambda_{0}\rangle
\exp\{i\vec{Q}\cdot(\vec{r}_{j}-\vec{r}_{j^{\prime}})\} \Biggr) \\ + {\rm B.G.},
\label{eq:xs_general}
\end{multline}
where $C_{2}$ is a scale factor of the inelastic scattering intensity, $\alpha$ and $\beta$ are the components of the Cartesian coordinate, and $\hat{S}$ is a spin operator~\cite{Marshall_1971,cmmnt_2}. The directional term $(\delta_{\alpha\beta} - Q_{\alpha}Q_{\beta}/|\vec{Q}|^2)$ can be considered as a constant by taking the orientational average over dynamically fluctuating molecules.

Figure~\ref{fig:models}(c) shows energy schemes obtained from Eqs.~(\ref{eq:H_hexa1}) and (\ref{eq:H_hexa2}) with antiferromagnetic $J_{1}=1.8$ meV and $J_{2}=2.4$ meV; these are in good agreement with the experimental resonance energies. Interestingly, the two spin-3/2 hexamer systems have a nonmagnetic singlet ground state ($S^{\rm tot}=0$) and triplet first-excited states ($S^{\rm tot}=1$), where $\vec{S}^{\rm tot}=\sum_{i=1}^{6}\vec{S}_{i}$. The states are represented by complex linear combinations of $S^{z}=\pm3/2$ and $\pm1/2$. A direct product of the hexa-I and hexa-II singlets generates a dodecamer-singlet ground state, and the two types of hexamer singlet-triplet excitations correspond to its intra-activations. A similar dodecamer singlet unit with six spin-1/2 dimers, called a pinwheel, was recently discovered in a kagome material by single-crystal inelastic neutron scattering~\cite{Matan_2010}.
The calculated $Q$ dependences of intensity for the two types of hexamer excitations, obtained from Eq.~(\ref{eq:xs_general}), are completely homothetic to those for the quasielastic excitations (broken and dotted curves in Fig.~\ref{fig:models}(a)), respectively, whose summation is again in good agreement with the experimental data.

For the hexa-I part, we also calculated the intensity distributions in the $hk$0 and $hhl$ zones from both Eqs.~(\ref{eq:xs_el}) and (\ref{eq:xs_general}) without powder orientational averaging. The patterns obtained from both these equations are the same, which is consistent with the {\it single-crystal} data measured below and above $T_N$ in MgCr$_2$O$_4$~\cite{Tomiyasu_2008}.
%dE_cal = 0.706

%
%\section{Discussion}
%
%Summary of the above analyses for non-experts of neutron scattering
%Origin of the hexa-I instability and hexa-II stability:
In this way, spin excitations above and below $T_N$ in HgCr$_2$O$_4$ were explained by the kagome-star dodecamer model with hexa-I and hexa-II, which is based on the proximity of $J_1$ and $J_2$. Then, the following question arises: what is the reason for this proximity in HgCr$_2$O$_4$?
In chromates, $J_1$ involves a direct exchange with a Cr-Cr direct overlap of $t_{2g}$ orbitals, and $J_2$ involves superexchange with a Cr-O-$A$-O-Cr path, with a Cr-O-$A$ bond angle of about $125^{\circ}$. The magnitude of $J_1$ weakens with increasing Cr-Cr distance, and that of $J_2$ strengthens with increasing bond angles up to 180$^{\circ}$ according to the Goodenough-Kanamori rules. Superexchange also occurs with 90$^{\circ}$ Cr-O-Cr paths, which can be neglected according to the rules. Meanwhile, as the ionic radius of an $A^{2+}$ cation increases from Mg$^{2+}$ to Hg$^{2+}$, the Cr-Cr distance increases from 2.94 {\AA} to 3.06 {\AA} and the Cr-O-$A$ angle increases from 122$^{\circ}$ to 131$^{\circ}$~\cite{Ehrenberg_2002,Weil_2006}. Thus, $J_1$ is expected to weaken and $J_2$ is expected to strengthen in HgCr$_2$O$_4$. This tendency is also supported by recent LSDA+$U$ calculations~\cite{Yaresko_2008}.

%the coexistence conjecture of quantum spin liquid and magnetic order
The above analyses revealed that the singlet-triplet picture, which is usually based on the quantum spin-1/2 dimer, can be extended to spin-3/2 hexamers and the dodecamer. Singlet formation does not imply complete disappearance of magnetic moments and can coexist with magnetic order, since the $g$ factor is arbitrary in the analyses. In fact, powder neutron diffraction experiments revealed that $A$Cr$_2$O$_4$ ($A = $ Mg, Zn, Hg) exhibits an ordered moment with only 2.2, 2.0, and 1.7 $\mu_{\rm B}$~\cite{Shaked_1970,Ji_2009,Matsuda_2007}. These values are around 1 $\mu_{\rm B}$ lower than the full value of 3 $\mu_{\rm B}$, which is consistent with the {\it partial} singlet formation conjecture. Thus, in the sense of the singlet, magnetically ordered phases in $A$Cr$_2$O$_4$ could be cited as a new class of quantum spin liquid with classical magnetic long-range order. Then, quasielastic spin excitations in a paramagnetic phase, whose the origin has been one of the longstanding issues related to frustration, could be interpreted as its thermally fluctuating spin-liquid state.

%
%\section{Conclusions}
%
In summary, we discovered almost degenerated spin-excitation modes in HgCr$_2$O$_4$ by powder inelastic neutron scattering. The $Q$ dependence of intensity demonstrates the coexistence of the normal hexa-I-type excitation and another type of excitation. On the basis of dynamical spin structure analyses, we propose the use of the kagome-star dodecamer model with hexa-I and hexa-II, which is based on the proximity between $J_1$ and $J_2$. Further, by numerically analyzing model Hamiltonians, we proposed a quantum-mechanical picture that is applicable to all chromates: the quantum spin liquid (spin-3/2 hexamer/dodecamer singlet) coexists with magnetic long-range order as ground states below $T_N$, and the quasielastic excitations correspond to its thermally fluctuating spin-liquid state above $T_N$. Further experimental and theoretical studies will be needed to prove the spin liquid conjecture.

\acknowledgments
We thank Mr. T. Asami for providing assistance at the JAEA, Mr. M. Onodera for providing assistance at Tohoku University, and Professors T. Masuda and T. J. Sato for fruitful discussions. The neutron experiments at the JAEA were performed under User Programs conducted by ISSP. This study was financially supported by Grants-in-Aid for Young Scientists (B) (22740209), Priority Areas (22014001), Scientific Researches (S) (21224008) and (A) (22244039), and Innovative Areas (20102005) from the MEXT of Japan; it was also supported by the Inter-university Cooperative Research Program of the Institute for Materials Research at Tohoku University.

\bibliography{HgCr2O4_5_EPL_PRL}

\begin{thebibliography}{28}
\bibitem[{cmm({\natexlab{b}})}]{cmmnt_2}
\bibinfo{note}{Equation~(\ref{eq:xs_el}) is obtained from
  Eq.~(\ref{eq:xs_general}) only when the matrix elements
  $\langle\lambda_{n}\mid\hat{S}_{j}^{\alpha}\mid\lambda_{0}\rangle$ can be
  simply reduced for elastic scattering.}
\end{thebibliography}

\end{document}